\documentclass[11pt,a4paper,reqno]{amsart}

\usepackage{times}
\usepackage{xspace}
\usepackage[inline]{enumitem}
\usepackage[numbers,sort&compress,longnamesfirst]{natbib}
\usepackage[hmargin=2.0cm,vmargin=2.8cm]{geometry}
\usepackage[colorlinks = true, linkcolor = blue, citecolor = blue]{hyperref}
\usepackage{amsfonts, comment}
\usepackage{amssymb, amsmath, latexsym, mathtools}
\usepackage{amsthm}
\usepackage[capitalize]{cleveref}
\usepackage{color, graphicx} 
\usepackage{todonotes}
\usepackage{lipsum}  
\usepackage{bbm}
\usepackage{algorithm}
\usepackage{algpseudocode}
\usepackage{tabu}
\usepackage{subcaption}

\newlist{myitems}{enumerate}{1}
\setlist[myitems]{label=\arabic*, font=\bfseries, resume}
\numberwithin{equation}{section}
\theoremstyle{plain}

\theoremstyle{definition}

\crefname{assumption}{Assumption}{Assumptions}
\Crefname{assumption}{Assumption}{Assumptions}

\def\<{\langle}
\def\>{\rangle}

\def\ud{\mathrm d}
\def\ue{\mathrm e}
\DeclareMathOperator*{\argmin}{arg\, min}

\newcommand{\cH}{\mathcal{H}}

\newcommand{\abs}[1]{\left|#1\right|} 

\begin{document}

\title[Error Analysis of Deep PDE Solvers for Option Pricing]{Error Analysis of Deep PDE Solvers for Option Pricing}
\author[J. Rou]{Jasper Rou}

\address{Delft Institute of Applied Mathematics, EEMCS, TU Delft, 2628CD Delft, The Netherlands}
\email{\href{mailto:j.g.rou@tudelft.nl}{j.g.rou@tudelft.nl}}



\begin{abstract}
Option pricing often requires solving partial differential equations (PDEs). Although deep learning-based PDE solvers have recently emerged as quick solutions to this problem, their empirical and quantitative accuracy remain not well understood, hindering their real-world applicability. In this research, our aim is to offer actionable insights into the utility of deep PDE solvers for practical option pricing implementation. Through comparative experiments in both the Black--Scholes and the Heston model, we assess the empirical performance of two neural network algorithms to solve PDEs: the Deep Galerkin Method and the Time Deep Gradient Flow method (TDGF). We determine their empirical convergence rates and training time as functions of (i) the number of sampling stages, (ii) the number of samples, (iii) the number of layers, and (iv) the number of nodes per layer. For the TDGF, we also consider the order of the discretization scheme and the number of time steps.
\end{abstract}

\maketitle

\section{Introduction}
Option pricing is a fundamental problem in finance. Since the seminal work of \citet{black1973pricing}, numerous mathematical models and computational approaches have been developed to determine option prices. One common approach formulates the price of an option as the solution to a partial differential equation (PDE), which can be solved numerically using methods such as finite differences or finite elements. However, traditional grid-based methods suffer from the curse of dimensionality: the number of grid points grows exponentially with the dimension of the problem. This challenge is particularly acute in high-dimensional settings, such as basket options or Markovian approximations of rough volatility models \cite{abi2019multifactor,papapantoleon2024time}.

Neural networks provide a promising alternative by efficiently approximating solutions to PDEs. Once trained, they can generate option prices rapidly, bypassing the limitations of conventional numerical methods. Deep learning-based approaches have been successfully applied in other financial contexts, such as risk management \cite{buehler2019deep}, portfolio optimization \cite{zhang2020deep}, and optimal stopping \cite{becker2019deep}, as well as data-driven methods to price options \cite{liu2019pricing}. Given their potential, various deep learning-based PDE solvers have been proposed \cite{gonon2024overview}, including Backward Stochastic Differential Equation (BSDE) methods \citep{han2018solving}, Deep Galerkin Methods (DGMs) \cite{sirignano2018dgm} and Time Deep Gradient Flow (TDGF) methods \cite{georgoulis2023discrete, papapantoleon2024time}. However, their empirical accuracy remains not sufficiently well understood, which limits their practical adoption in financial applications.

This study focuses on two neural network methods for solving PDEs: the DGM and the TDGF. Our primary objective is to assess their empirical accuracy. For theoretical convergence analyses, see the work of \citet{jiang2023global} for DGM and \citet{liu2025convergence} for TDGF. For empirical studies of other deep PDE solvers, such as BSDE-based methods, see the work of \citet{assabumrungrat2024error}.

To provide actionable insights into the applicability of deep PDE solvers for option pricing, we systematically analyze the impact of key parameters on accuracy and training time. First, we investigate the effect of training by varying the number of sampling stages and the number of samples. Second, we investigate the effect of the size of the neural network by varying the number of layers and the number of nodes per layer. For a comparison of different architectures, see the work of \citet{van2023machine}. Finally, for TDGF, we also examine the discretization order and the number of time steps.

Our main findings are: the $L^2$-error decreases almost linearly with the number of sampling stages; the number of layers tends to decrease the error, but not with a clear rate; and increasing the number of time steps decreases the $L^2$-error with the second-order method decreasing quicker than the first-order method. These three parameters increase the training time linearly. The number of samples and nodes per layer did not show a clear relationship with neither the $L^2$-error nor the training time. The first three parameters concern training stages done sequentially, while the other two concern training stages which can be done in parallel.

The remainder of this paper is structured as follows. \Cref{sec:neural} introduces the two neural network-based PDE solvers. \Cref{sec:option} outlines the option pricing models under consideration: Black–Scholes and Heston. \Cref{sec:implementation} details the implementation aspects. \Cref{sec:results} presents the numerical results for each of the five parameters. Finally, \Cref{sec:conclusion} summarizes our findings and conclusions.

\section{Neural network methods}
\label{sec:neural}
This section explains the two neural network methods used in this paper. Subsection \ref{sec:time} elaborates on the TDGF and Subsection \ref{sec:deep} on the DGM.

\subsection{Time Deep Gradient Flow Method}
\label{sec:time}
The TDGF is a neural network method to efficiently solve high-dimensional PDEs \cite{papapantoleon2024time, georgoulis2023discrete, georgoulis2024deep}. Consider the general PDE
\[
\begin{aligned}
    \frac{\partial}{\partial t} u(t, \mathbf{x}) + \mathcal{A} u(t, \mathbf{x}) + ru(t, \mathbf{x}) = 0, \quad & (t, \mathbf{x}) \in [0,T] \times \Omega,\\
    u(0, \mathbf{x}) = \Psi(\mathbf{x}), \quad & \mathbf{x} \in \Omega,
\end{aligned}
\]
with $\mathcal{A}$ a second-order differential operator of the form
\begin{equation}
\label{eq:generator}
    \mathcal{A} u = - \sum_{i,j=1}^d a^{ij} \frac{\partial^2 u}{\partial x_i \partial x_j} + \sum_{i=1}^{d} \beta^i \frac{\partial u}{\partial x_i}.    
\end{equation}
Using the splitting method from \citet{papapantoleon2024time}, $\mathcal{A}$ can be rewritten in the form
\begin{equation}
\label{eq:operator_form}
    \mathcal{A} u = - \nabla \cdot \left( A \nabla u \right) + \mathbf{b} \cdot \nabla u,   
\end{equation}
with a symmetric an positive semi-definite matrix
\begin{equation}
\label{eq:coefficients}
A = \begin{bmatrix}
    a^{11} & a^{21} & \dots & a^{d1} \\
    a^{21} & a^{22} & \dots & a^{d2} \\
    \vdots & \vdots & \ddots & \vdots \\
    a^{1d} & a^{2d} & \dots & a^{dd}
\end{bmatrix}
\quad \textrm{and vector } \quad
\mathbf{b} = \begin{bmatrix}
    b^1 \\
    b^2 \\
    \vdots \\
    b^d
\end{bmatrix}.
\end{equation}
Let us divide the time interval $(0,T]$ into $K$ equally spaced intervals $(t_{k-1},t_k]$, with $h = t_k - t_{k-1} = \frac{1}{K}$ for $k=0,1,\dots,K$.
Let $U^k$ denote the approximation to the solution of the PDE $u(t_k, \mathbf{x})$ at time step $t_k$, using either a first- or second-order discretization scheme \cite{akrivis2004implicit}
\[
\begin{aligned}
\frac{U^k - U^{k-1}}{h} - \nabla \cdot \left( A \nabla U^k \right) + F \left( U^{k-1} \right) + r U^k & = 0, \\
\frac{\frac{3}{2} U^k - 2 U^{k-2} + \frac{1}{2} U^{k-1}}{h} - \nabla \cdot \left( A \nabla U^k \right) + 2 F \left( U^{k-1} \right) - F \left( U^{k-2} \right) + r U^k & = 0,
\end{aligned}
\]
with $F(u) = \mathbf{b} \cdot \nabla u$, $U^0 = \Psi$ and in the second-order scheme we take $U^1$ from the first-order scheme.
Then we can rewrite the discretized PDE as an energy functional \cite{papapantoleon2024time, georgoulis2024deep}
\[
U^k = \argmin_{u \in \cH_0^1} I^k (u),
\]
with $ \cH_0^1$ the Sobolev space in which the derivatives up to order 1 have finite $L^2$-norm and the energy functionals
\begin{equation}
\label{eq:energy}
\begin{aligned}
I_1^k(u) = & \frac{1}{2} \left \Vert u - U^{k-1} \right \Vert_{L^2(\Omega)}^2 + h \left( \int_{\Omega} \frac{1}{2} \left( \left( \nabla u \right)^\mathsf{T} A \nabla u + r u^2 \right) + F \left( U^{k-1} \right) u \ud x \right), \\
I_2^k(u) = & \frac{1}{2} \left \Vert u - \frac{4}{3} U^{k-1} + \frac{1}{3} U^{k-2} \right \Vert_{L^2(\Omega)}^2 \\
& + \frac{2h}{3} \left( \int_{\Omega} \frac{1}{2} \left( \left( \nabla u \right)^\mathsf{T} A \nabla u + r u^2 \right) + \left( 2 F \left( U^{k-1} \right) - F \left( U^{k-2} \right) \right) u \ud x \right),
\end{aligned}
\end{equation}
for the first- and second-order discretization respectively.
Let $f^k(\mathbf{x}; \theta)$ denote a neural network approximation of $U^k$ with trainable parameters $\theta$. 
Applying a Monte Carlo approximation to the integrals, the discretized cost functional takes the form
\begin{equation}
\label{eq:disctretized_loss}
L_n^k \left( \theta ; \mathbf{x} \right) = \frac{|\Omega|}{2M} \sum_{m=1}^{M} \left( f^k(\mathbf{x}_m;\theta) + \sum_{j=1}^n \alpha_n^j f^{k-j}(\mathbf{x}_m) \right)^2 + \beta_n h N_n^k \left( \theta ; \mathbf{x} \right),
\end{equation}
with
\begin{align*}
N_n^k \left( \theta ; \mathbf{x} \right) 
    = & \frac{|\Omega|}{M} \sum_{m=1}^{M} \Bigg[ \frac{1}{2} \left( \left( \nabla f^k(\mathbf{x}_m; \theta) \right)^\mathsf{T} A \nabla f^k(\mathbf{x}_m; \theta) + r \left( f^k({\mathbf{x}_m}; \theta) \right) ^2 \right) \\
    & + \left( \mathbf{b} \cdot \sum_{j=1}^n \gamma_n^j \nabla f^{k-j}(\mathbf{x}_m) \right) f^k(\mathbf{x}_m; \theta) \Bigg].
\end{align*}
Here, $M$ denotes the number of samples $\mathbf{x}_m$; $n \in \{1,2\}$ the order of the discretization and $\alpha_n^j, \beta_n$ and $\gamma_n^j$ are the corresponding coefficients.

In order to minimize this cost function, we use a stochastic gradient descent-type algorithm, \textit{i.e.} an iterative scheme of the form: 
\begin{equation}
\label{eq:SGD}
    \theta_{n+1} = \theta_n - \alpha \nabla_{\theta} L^k(\theta_n; \mathbf{x}).
\end{equation}
The hyperparameter $\alpha$ is the step size of our update, called the learning rate. 
An overview of the TDGF method appears in Algorithm \ref{alg:TDGF}.

\begin{algorithm}
\caption{Time Deep Gradient Flow method}
\label{alg:TDGF}
\begin{algorithmic}[1]
\State Initialize $\theta_0^0$.
\State Set $f^0(\mathbf{x};\theta) = \Psi(\mathbf{x})$.
\For{each time step $k = 1,\dots,K$}
\State Initialize $\theta_0^k = \theta^{k-1}$.
\For{each sampling stage $n=1,...,N$}
\State Generate $M$ random points $\mathbf{x}_m$ for training.
\State Calculate the cost functional $L^k(\theta_n^k; \mathbf{x})$ for the selected points.
\State Take a descent step $\theta_{n+1}^k = \theta_n^k - \alpha \nabla_{\theta} L^k(\theta_n^k; \mathbf{x})$.
\EndFor
\EndFor
\end{algorithmic}
\end{algorithm}

\subsection{Deep Galerkin Method}
\label{sec:deep}
We compare the TDGF method with a popular deep learning method for solving PDEs: the DGM of \citet{sirignano2018dgm}. 
In the DGM approach, we minimize the square $L^2$-error of the PDE: 
\[
\left \Vert \frac{\partial u}{\partial t} - \nabla \cdot \left( A \nabla u \right) + \mathbf{b} \cdot \nabla u + ru \right \Vert_{L^2([0,T] \times \Omega)}^2 
+ \left \Vert u(0,x) - \Psi(x) \right \Vert_{L^2(\Omega)}^2. 
\]
Then the cost functional for the neural network approximation $f(t,\mathbf{x}; \theta)$ of $u$, takes the form 
\[
\begin{aligned}
L(\theta; t, \mathbf{x}) = & \frac{T \abs{\Omega}}{M_1} \sum_{m=1}^{M_1}
\left[ f(t, \mathbf{x}_m; \theta) - \nabla \cdot \left( A \nabla f(t, \mathbf{x}_m; \theta) \right) + \mathbf{b} \cdot \nabla f(t, \mathbf{x}_m; \theta) + r f(t, \mathbf{x}_m; \theta) \right] \\
& + \frac{\abs{\Omega}}{M_2} \sum_{m=1}^{M_2} \left[ f(0, \mathbf{x}_m; \theta) - \Psi(\mathbf{x}_m) \right]. 
\end{aligned}
\]
The solution of the PDE is approximated by a neural network using stochastic gradient descent as in equation \eqref{eq:SGD}.
Contrary to the TDGF there is no time stepping. Instead of training a neural network for each time step, there is one neural network with time as input parameter. An overview of the DGM appears in Algorithm \ref{alg:DGM}.

\begin{algorithm}
\caption{Deep Galerkin Method}
\label{alg:DGM}
\begin{algorithmic}[1]
\State Initialize $\theta_0$.
\For{each sampling stage $n=1,...,N$}
\State Generate $M$ random points $(t_m, \mathbf{x}_m)$ for training.
\State Calculate the cost functional $L(\theta_n; t, \mathbf{x})$ for the selected points.
\State Take a descent step $\theta_{n+1} = \theta_n - \alpha \nabla_{\theta} L(\theta_n; t, \mathbf{x})$.
\EndFor
\end{algorithmic}
\end{algorithm}

\section{Option pricing models}
\label{sec:option}
This section explains the two option pricing models in which we solve the pricing PDE. Subsection \ref{sec:BS} elaborates on the Black--Scholes model and Subsection \ref{sec:Heston} on the Heston model.

\subsection{Black--Scholes}
\label{sec:BS}
In the \citet{black1973pricing} model, the dynamics of the stock price $S$ is a geometric Brownian motion:
\[
    \ud S_t = r S_t \ud t + \sigma S_t \ud W_t, \quad S_0 > 0,
\]
with $r, \sigma \in \mathbb{R}_+$ the risk free rate and the volatility respectively. 

Consider a European call option on $S$ with payoff $\Psi(S_T) = \left(S_T - K \right)^+$ at maturity time $T>0$.
Using the fundamental theorem of asset pricing and the Feynman--Kac formula, the price of this derivative can be written as the solution to a PDE in this model. 
Indeed, let $u: [0,T] \times \Omega \to \mathbb{R}$ denote the price of this derivative, with $\Omega \subseteq \mathbb{R}$ and $t$ the time to maturity. 
Then, $u$ solves the Black--Scholes PDE:
\[
\begin{aligned}
    \frac{\partial u}{\partial t} -\frac{1}{2} \sigma^2 S^2 \frac{\partial^2 u}{\partial S^2} - r S \frac{\partial u}{\partial S} + r u & = 0, \quad && (t,S) \in [0,T] \times \Omega,\\
    u(0,S) & = \Psi(S), \quad && S \in \Omega.
\end{aligned}
\]
This PDE has an exact solution:
\[
u(t,S) = S \Phi(d_1) - K \ue^{-r t} \Phi(d_2),
\]
with $\Phi$ the standard normal cumulative distribution function, 
\[
d_1 = \frac{\log \left( \frac{S}{K} \right) + \left( r + \frac{\sigma^2}{2} \right) t}{\sigma \sqrt{t}} \quad \textrm{and} \quad d_2 = d_1 - \sigma \sqrt{t}.
\]
The operator $\mathcal{A}$ takes the form \eqref{eq:operator_form} with the coefficients in \eqref{eq:coefficients} provided by 
\begin{align*}
    a &= \frac{1}{2} \sigma^2 S^2, \\
    b &= (\sigma^2 - r) S.
\end{align*}

\subsection{Heston}
\label{sec:Heston}
The \citet{heston1993closed} model is a popular stochastic volatility model with dynamics
\[
\begin{aligned}
\ud S_t & = r S_t \ud t + \sqrt{V_t} S_t \ud W_t, \quad & S_0 > 0, \\
\ud V_t & = \lambda ( \kappa - V_t) \ud t + \eta \sqrt{V_t} \ud B_t, \quad & V_0 > 0.
\end{aligned}
\]
Here $V$ is the variance process, $W, B$ are correlated (standard) Brownian motions, with correlation coefficient $\rho$, and $\lambda, \kappa, \eta \in \mathbb{R}_+$. 
The generator corresponding to these dynamics, in the form \eqref{eq:generator}, equals
\[
\mathcal{A} u = - r S \frac{\partial u}{\partial S} - \lambda (\kappa - V) \frac{\partial u}{\partial V} - \frac{1}{2} S^2 V \frac{\partial^2 u}{\partial S^2} - \frac{1}{2} \eta^2 V \frac{\partial^2 u}{\partial V^2} - \rho \eta S V \frac{\partial^2 u}{\partial S \partial V}.
\]
This PDE does not have an exact solution. The characteristic function of the Heston model does have an analytical representation \cite{heston1993closed}, from which a reference price can be determined using the COS method \cite{fang2009novel}.

The operator $\mathcal{A}$ takes the form \eqref{eq:operator_form} with the coefficients in \eqref{eq:coefficients} provided by 
\begin{align*}
    a^{11} &= \frac{1}{2} S^2 V, \\
    a^{21} = a^{12} &= \frac{1}{2} \rho \eta S V, \\
    a^{22} & = \frac{1}{2} \eta^2 V, \\
    b^1 &= \left(-r + V + \frac{1}{2} \rho \eta \right) S, \\
    b^2 &= \lambda (V - \kappa) + \frac{1}{2} \eta^2 + \frac{1}{2} \rho \eta V.
\end{align*} 

\section{Implementation details}
\label{sec:implementation}
For the architecture we use the architecture from \citet{papapantoleon2024time} including the use of information about the option price in order to facilitate the training of the neural network:
\begin{align*}
    X^1 & = \sigma_1 \left( W^1 \mathbf{x} + b^1 \right),  \\
    Z^l & = \sigma_1 \left( U^{z,l} \mathbf{x} + W^{z,l} X^l + b^{z,l} \right), \quad & l=1,\dots,L, \\
    G^l & = \sigma_1 \left( U^{g,l} \mathbf{x} + W^{g,l} X^l + b^{g,l} \right), \quad & l=1,\dots,L, \\
    R^l & = \sigma_1 \left( U^{r,l} \mathbf{x} + W^{r,l} X^l + b^{r,l} \right), \quad & l=1,\dots,L, \\
    H^l & = \sigma_1 \left( U^{h,l} \mathbf{x} + W^{h,l} \left( X^l \odot R^l \right) + b^{h,l} \right), \quad & l=1,\dots,L, \\
    X^{l+1} & = \left( 1 - G^l \right) \odot H^l + Z^l \odot X^l, \quad & l=1,\dots,L, \\
    f(\mathbf{x};\theta) & = \left(S - K \ue^{-rt} \right)^+ + \sigma_2 \left( W X^{L+1} + b \right), 
\end{align*}
with activation functions the hyperbolic tangent function, $\sigma_1(x) = \tanh(x)$, and the softplus function, $\sigma_2(x) = \log \left( \ue^x +1 \right)$, which guarantees that the option price remains above the no-arbitrage bound. The parameters of the network have dimensions $W^1, U^{m,l} \in \mathbb{R}^{D \times d}$; $b^1, b^{m,l} \in \mathbb{R}^D$; $W^{m,l} \in \mathbb{R}^{D \times D}$; $W \in \mathbb{R}^{1 \times D}$ and $b \in \mathbb{R}$ for $m=z,g,r,h$ and $l=1,...,L$, with $\mathbf{x} \in \mathbb{R}^d$.

We consider the effect of five parameters on the error: the number of sampling stages, $N$ in \cref{alg:TDGF}; the number of samples, $M$ in equation \eqref{eq:disctretized_loss}; the number of layers $L$; the number of nodes per layer $D$; and for the TDGF also the number of time steps, $K$ in Algorithm \ref{alg:TDGF}. In the last case we consider both the first- and second-order discretization scheme in equation \eqref{eq:energy}.
As the default parameter set we take 600 samples per dimension in each sampling stage. To obtain the total number of samples from this number, multiply by 2 for DGM in the Black--Scholes (time and stock price) and TDGF in the Heston model (stock price and volatility) and multiply by 3 for the DGM in the Heston model (time, stock price and volatility). The default network size is 3 layers and 50 nodes per layers. For the TDGF we take 100 time steps and 500 sampling stages in each time step and for the DGM we take 100,000 sampling stages. After this many sampling stages the error does not decrease further.

For both DGM and TDGF we use the Adam optimizer \cite{kingma2014adam} with a learning rate $\alpha = 3 \times 10^{-4}$, $(\beta_1,\beta_2) = (0.9,0.999)$ and zero weight decay.
The training is performed on the DelftBlue supercomputer \cite{DHPC2022}, using one seventh instance of a NVidia Tesla A100 GPU.
We run each problem for five different random seed and compare the average error of the five runs.

\begin{figure}
    \centering
    \includegraphics[width=\linewidth]{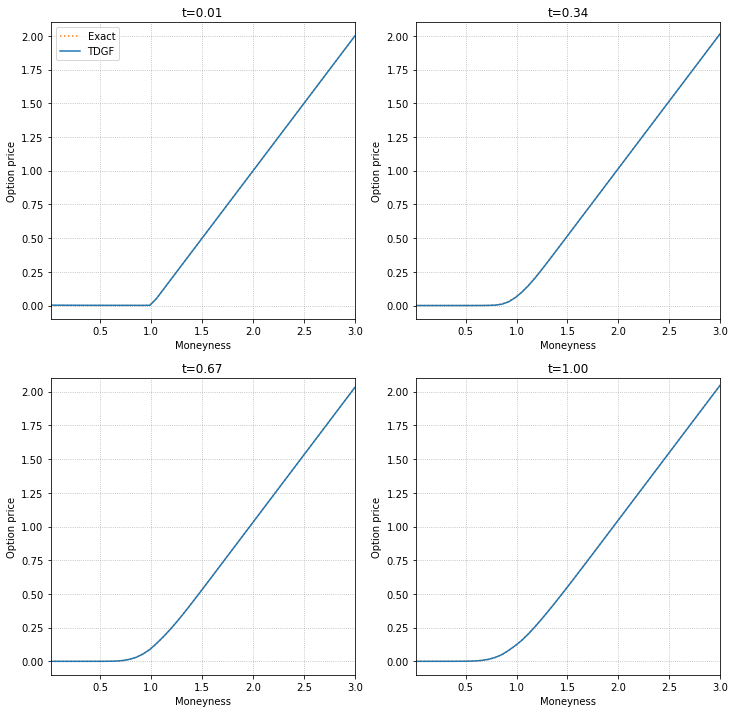}
    \caption{Exact price of European call option with $r=0.05$, $\sigma=0.25$ and $T=1.0$ compared to the price of computed by the TDGF.}
    \label{fig:BS_call}
\end{figure}

As the modeling problem we take the price of a European call option with interest rate $r=0.05$ and maturity $T=1.0$ year. We consider the Black--Scholes model with volatility $\sigma=0.25$ and the Heston model with $\eta=0.1$, $\rho=0.0$, $\kappa=0.01$ and $\lambda=2.0$.
For the domain $\Omega$ we consider $S \in [0.01,3.0]$ and $V \in [0.001,0.1]$.
The solution of the Black--Scholes PDE with these parameters together with the solution produced by the TDGF with the default training parameters is in \cref{fig:BS_call}.

\section{Results}
\label{sec:results}
In the next subsections we vary one of the parameters while keeping the others constant at the default value. We compute the $L^2$-error on an equidistant grid of 47 points in each dimension on the domain.

\subsection{Sampling stages}
First, we consider the number of sampling stages. For the TDGF we vary the number of sampling stages per time step from 16 to 500. For the DGM we vary the number of sampling stages from 2048 to 100,000. After this many sampling stages, the error does not decrease anymore in the Black--Scholes model. In the Heston model, the error seems to quit decreasing quicker for both methods. 
The fitted convergence rates for both methods and both models are in \cref{tab:sampling_stages}. All convergence rates are slightly larger than -1.
The plots of the $L^2$-error on linear and log scale are in Figures \ref{fig:sampling_stages_BS_TDGF}-\ref{fig:sampling_stages_Heston_DGM} together with the training time. The training time increases linearly with the number of sampling stages. 

\begin{table}
    \centering
    \begin{tabular}{l|r|r}
        Method & Black--Scholes & Heston \\ \hline
        TDGF & -0.91 & -0.63 \\ \hline 
        DGM & -0.73 & -0.75
    \end{tabular}
    \caption{Convergence rates for the number of sampling stages.}
    \label{tab:sampling_stages}
\end{table}

\subsection{Samples}
Second, we consider the number of samples per dimension in each sampling stage. We vary the number of samples per dimension from 16 to 600. The fitted convergence rates for both methods and both models are in \cref{tab:number_of_samples}. In general, it is hard to draw conclusions. For the TDGF the rates are slightly negative, but far from -0.5, which would be the expected rate of convergence for Monte Carlo sampling. For the DGM the rates are larger and the error does not decrease as uniformly with the number of samples as for the TDGF. The plots of the $L^2$-error on linear and log scale are in Figures \ref{fig:nSim_BS_TDGF}-\ref{fig:nSim_Heston_DGM} together with the training time. The number of samples does not have a big impact on the training time.

\begin{table}
    \centering
    \begin{tabular}{l|r|r}
        Method & Black--Scholes & Heston \\ \hline
        TDGF & -0.27 & -0.11 \\ \hline 
        DGM & -0.12 & 0.2
    \end{tabular}
    \caption{Convergence rates for the number of samples.}
    \label{tab:number_of_samples}
\end{table}

\subsection{Layers}
Third, we consider the number of layers of the neural network. We vary the number for layers from 1 to 4. The fitted convergence rates for both methods and both models are in \cref{tab:layers}. The rates vary but are all negative so, in general, more layers improves the result. The plots of the $L^2$-error on linear and log scale are in Figures \ref{fig:n_layers_BS_TDGF}-\ref{fig:n_layers_Heston_DGM} together with the training time. The training time increases linearly with the number of layers.

\begin{table}
    \centering
    \begin{tabular}{l|r|r}
        Method & Black--Scholes & Heston \\ \hline
        TDGF & -0.63 & -0.47 \\ \hline 
        DGM & -0.33 & -0.70
    \end{tabular}
    \caption{Convergence rates for the number of layers.}
    \label{tab:layers}
\end{table}

\subsection{Nodes per layer}
Fourth, we consider the number of nodes per layers of the neural network. We vary the number of layers from 10 to 50. The fitted convergence rates for both methods and both models are in \cref{tab:nodes_per_layer}. The rates vary across different methods and models and are even positive for the DGM. The plots of the $L^2$-error on linear and log scale are in Figures \ref{fig:nodes_per_layer_BS_TDGF}-\ref{fig:nodes_per_layer_Heston_DGM} together with the training time. The number of nodes per layers does not have a big impact on the training time.

\begin{table}
    \centering
    \begin{tabular}{l|r|r}
        Method & Black--Scholes & Heston \\ \hline
        TDGF & -1.11 & -0.39 \\ \hline 
        DGM & 0.15 & 0.07
    \end{tabular}
    \caption{Convergence rates for the number of nodes per layer.}
    \label{tab:nodes_per_layer}
\end{table}

\subsection{Time steps}
Fifth and final, we consider the number of time steps. We vary the number of time steps from 2 to 25. The fitted convergence rates for both models and for both first and second order time-stepping are in \cref{tab:time_steps}. After 25 time steps, the second-order scheme does not improve any further, but the first-order scheme does. The rates for the Black--Scholes are lower than for Heston. In both cases $O(2)$ outperforms $O(1)$. The plots of the $L^2$-error on linear and log scale are in Figures \ref{fig:N_t_BS_TDGF}-\ref{fig:N_t_Heston_TDGF} together with the training time. The training time grows linearly with the number of time steps with the second order method growing faster than the first order method.

\begin{table}
    \centering
    \begin{tabular}{l|r|r}
        Method & Black--Scholes & Heston \\ \hline
        $O(1)$ & -0.29 & -0.15 \\ \hline 
        $O(2)$ & -0.56 & -0.25  
    \end{tabular}
    \caption{Convergence rates for the number of time steps}
    \label{tab:time_steps}
\end{table}

\section{Conclusion}
\label{sec:conclusion}
This research analyzed the error of two neural network methods to solve option pricing PDEs: TDGF and DGM. We determined the empirical convergence rates of the $L^2$-error of five parameters in both the Black--Scholes and the Heston model. We also considered the effect of these parameters on the training time. Based on the experiments we can give some recommendations that can assist anyone who want to use the methods in a practical setting.
\begin{itemize}
    \item For both the TDGF and the DGM, the $L^2$-error decreases almost linearly with the number of sampling stages up to some point where it stops converging. Since the training time grows linearly with the the number of sampling stages, it would be optimal to stop at this point. Unfortunately, there is no method to locate this point beforehand and we recommend choosing the number of sampling stages based on whether speed or accuracy is more important in the practical setting.
    \item For the TDGF the $L^2$-error decreases slightly with the number of samples. Since the number of samples does not influence the training time, we recommend using a large number of samples like six hundred per dimension or even more.
    \item For the DGM the $L^2$-error does not decrease with the number of samples. Therefore, it is hard to give any recommendation.
    \item For both TDGF and DGM, the number of layers tends to decrease the error, but not with a clear rate. One layer is clearly not enough, but four layers does not improve the results compared to two or three layers. Since the number of layers has a big influence on the training time, we recommend using two or three layers.
    \item For the TDGF, the $L^2$-error decreases with the number of nodes per layer. Since the number of nodes per layer does not influence the training time, we recommend choosing a large number of nodes per layer like forty of fifty.
    \item For the DGM, the $L^2$-error does not decrease with the number of nodes per layer. We recommend choosing a smaller number of nodes per layer like thirty.
    \item For the TDGF, the number of time steps decreases the $L^2$-error. Using a second-order time-stepping method, the error decreases quicker than using a first-order method. We recommend using the second-order time stepping method. Since the training time increases linearly with the number of time steps, we again recommend choosing the number of time steps based on whether speed or accuracy is more important in the practical setting.
\end{itemize}

\bibliographystyle{abbrvnat}
\bibliography{references}

\appendix

\begin{figure}
    \centering
    \begin{subfigure}{0.49\textwidth}
        \centering
        \includegraphics[width=\linewidth]{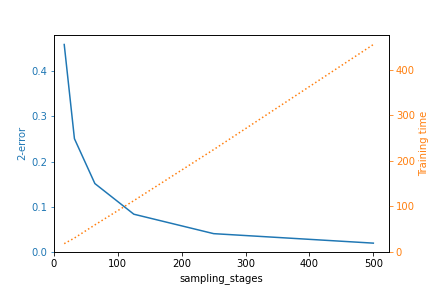}
        \caption{Linear scale}
    \end{subfigure}
    \begin{subfigure}{0.49\textwidth}
        \centering
        \includegraphics[width=\linewidth]{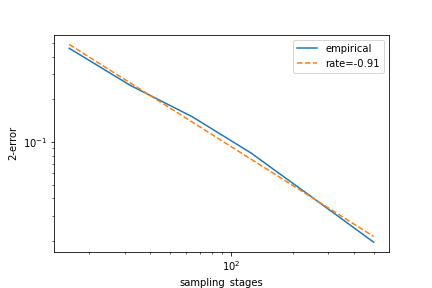}
        \caption{Logarithmic scale}
    \end{subfigure}
    \caption{$L^2$-error of the TDGF for a call option in the Black--Scholes model against number of sampling stages varying from 16 to 500.}
    \label{fig:sampling_stages_BS_TDGF}
\end{figure}

\begin{figure}
    \centering
    \begin{subfigure}{0.49\textwidth}
        \centering
        \includegraphics[width=\linewidth]{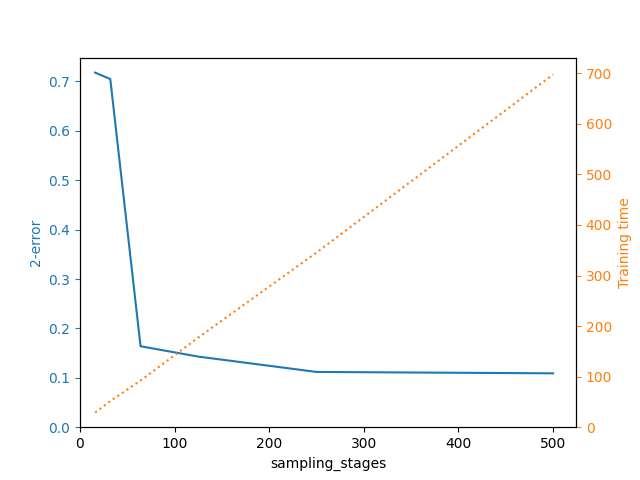}
        \caption{Linear scale}
    \end{subfigure}
    \begin{subfigure}{0.49\textwidth}
        \centering
        \includegraphics[width=\linewidth]{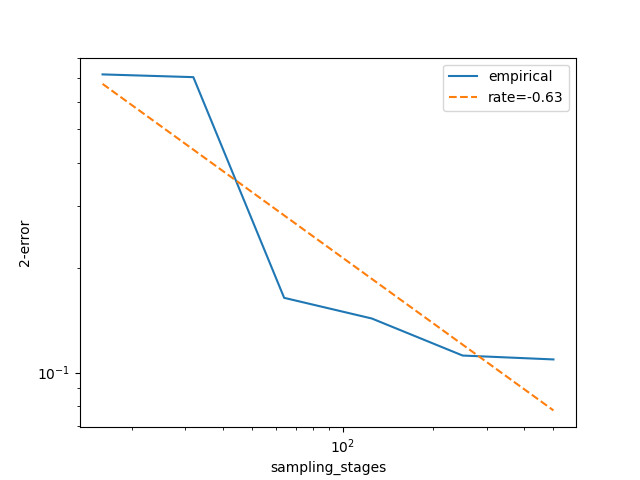}
        \caption{Logarithmic scale}
    \end{subfigure}
    \caption{$L^2$-error of the TDGF for a call option in the Heston model against number of sampling stages varying from 16 to 500.}
    \label{fig:sampling_stages_Heston_TDGF}
\end{figure}

\begin{figure}
    \centering
    \begin{subfigure}{0.49\textwidth}
        \centering
        \includegraphics[width=\linewidth]{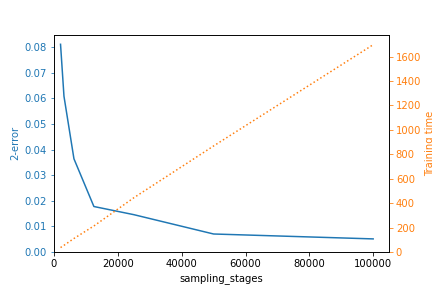}
        \caption{Linear scale}
    \end{subfigure}
    \begin{subfigure}{0.49\textwidth}
        \centering
        \includegraphics[width=\linewidth]{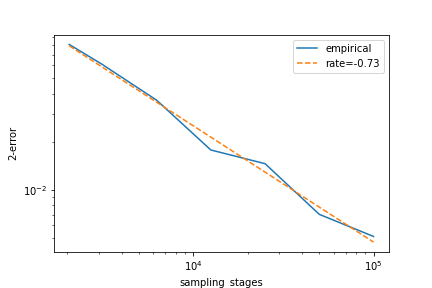}
        \caption{Logarithmic scale}
    \end{subfigure}
    \caption{$L^2$-error of the DGM for a call option in the Black--Scholes model against number of sampling stages varying from 2048 to 100,000.}
    \label{fig:sampling_stages_BS_DGM}
\end{figure}

\begin{figure}
    \centering
    \begin{subfigure}{0.49\textwidth}
        \centering
        \includegraphics[width=\linewidth]{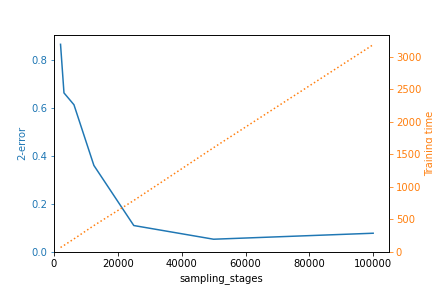}
        \caption{Linear scale}
    \end{subfigure}
    \begin{subfigure}{0.49\textwidth}
        \centering
        \includegraphics[width=\linewidth]{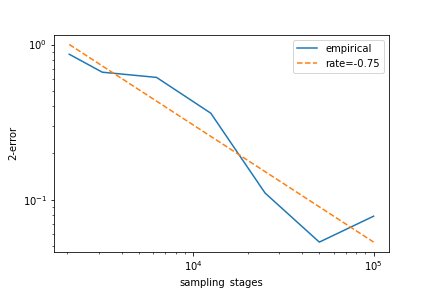}
        \caption{Logarithmic scale}
    \end{subfigure}
    \caption{$L^2$-error of the DGM for a call option in the Heston model against number of sampling stages varying from 2048 to 100,000.}
    \label{fig:sampling_stages_Heston_DGM}
\end{figure}

\begin{figure}
    \centering
    \begin{subfigure}{0.49\textwidth}
        \centering
        \includegraphics[width=\linewidth]{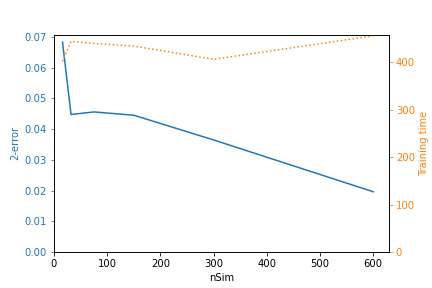}
        \caption{Linear scale}
    \end{subfigure}
    \begin{subfigure}{0.49\textwidth}
        \centering
        \includegraphics[width=\linewidth]{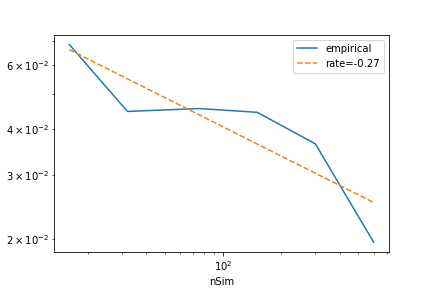}
        \caption{Logarithmic scale}
    \end{subfigure}
    \caption{$L^2$-error of the TDGF for a call option in the Black--Scholes model against number of samples varying from 16 to 600.}
    \label{fig:nSim_BS_TDGF}
\end{figure}

\begin{figure}
    \centering
    \begin{subfigure}{0.49\textwidth}
        \centering
        \includegraphics[width=\linewidth]{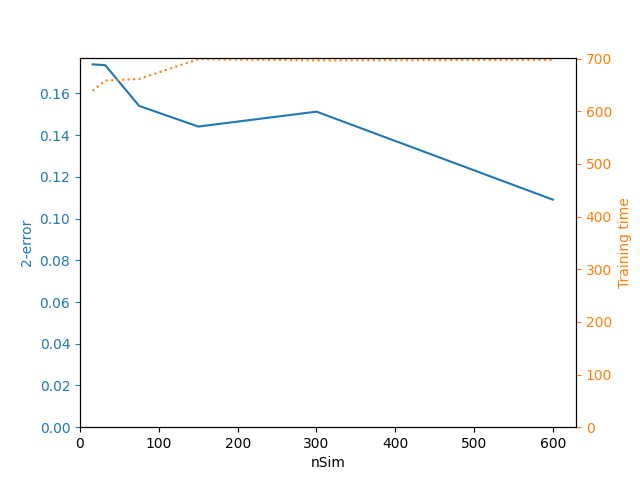}
        \caption{Linear scale}
    \end{subfigure}
    \begin{subfigure}{0.49\textwidth}
        \centering
        \includegraphics[width=\linewidth]{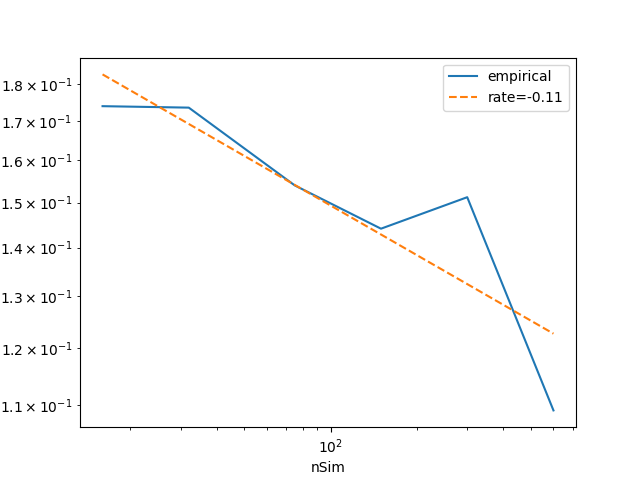}
        \caption{Logarithmic scale}
    \end{subfigure}
    \caption{$L^2$-error of the TDGF for a call option in the Heston model against number of samples varying from 16 to 600.}
    \label{fig:nSim_Heston_TDGF}
\end{figure}

\begin{figure}
    \centering
    \begin{subfigure}{0.49\textwidth}
        \centering
        \includegraphics[width=\linewidth]{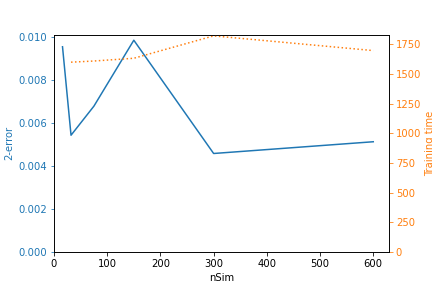}
        \caption{Linear scale}
    \end{subfigure}
    \begin{subfigure}{0.49\textwidth}
        \centering
        \includegraphics[width=\linewidth]{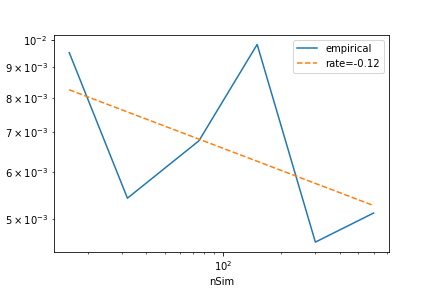}
        \caption{Logarithmic scale}
    \end{subfigure}
    \caption{$L^2$-error of the DGM for a call option in the Black--Scholes model against number of samples varying from 16 to 600.}
    \label{fig:nSim_BS_DGM}
\end{figure}

\begin{figure}
    \centering
    \begin{subfigure}{0.49\textwidth}
        \centering
        \includegraphics[width=\linewidth]{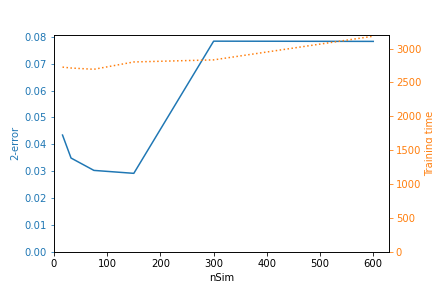}
        \caption{Linear scale}
    \end{subfigure}
    \begin{subfigure}{0.49\textwidth}
        \centering
        \includegraphics[width=\linewidth]{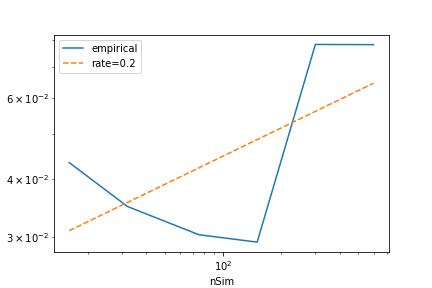}
        \caption{Logarithmic scale}
    \end{subfigure}
    \caption{$L^2$-error of the DGM for a call option in the Heston model against number of samples varying from 16 to 600.}
    \label{fig:nSim_Heston_DGM}
\end{figure}

\begin{figure}
    \centering
    \begin{subfigure}{0.49\textwidth}
        \centering
        \includegraphics[width=\linewidth]{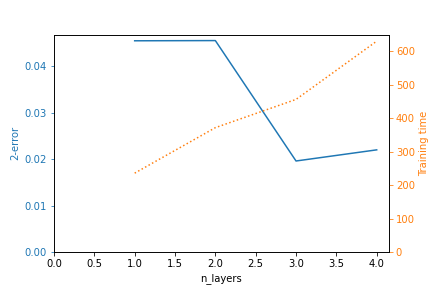}
        \caption{Linear scale}
    \end{subfigure}
    \begin{subfigure}{0.49\textwidth}
        \centering
        \includegraphics[width=\linewidth]{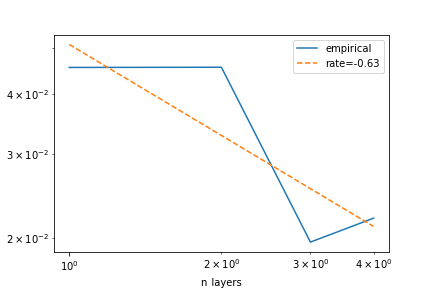}
        \caption{Logarithmic scale}
    \end{subfigure}
    \caption{$L^2$-error of the TDGF for a call option in the Black--Scholes model against number of layers varying from 1 to 4.}
    \label{fig:n_layers_BS_TDGF}
\end{figure}

\begin{figure}
    \centering
    \begin{subfigure}{0.49\textwidth}
        \centering
        \includegraphics[width=\linewidth]{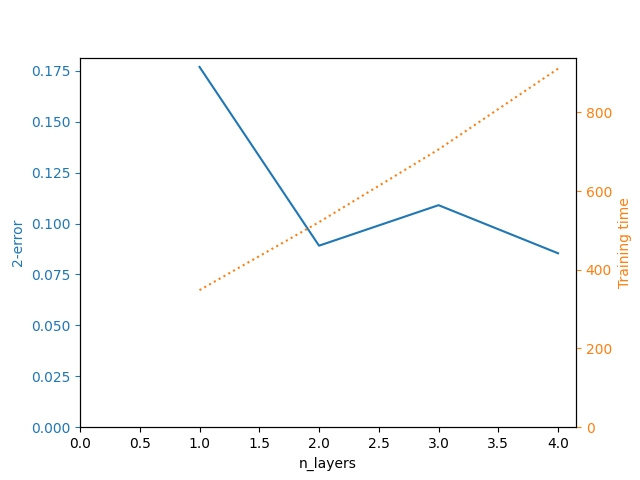}
        \caption{Linear scale}
    \end{subfigure}
    \begin{subfigure}{0.49\textwidth}
        \centering
        \includegraphics[width=\linewidth]{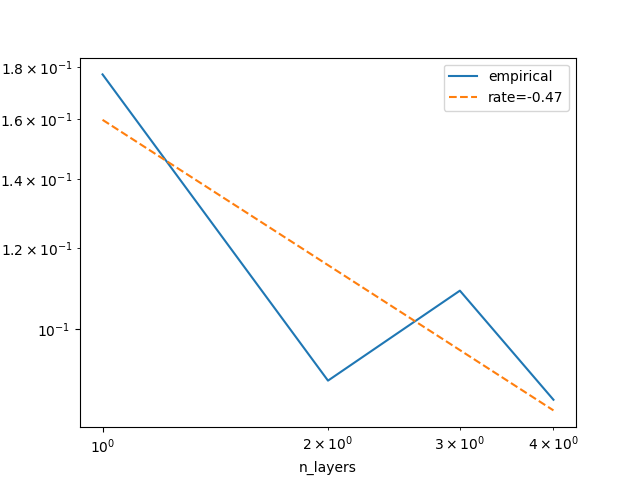}
        \caption{Logarithmic scale}
    \end{subfigure}
    \caption{$L^2$-error of the TDGF for a call option in the Heston model against number of layers varying from 1 to 4.}
    \label{fig:n_layers_Heston_TDGF}
\end{figure}

\begin{figure}
    \centering
    \begin{subfigure}{0.49\textwidth}
        \centering
        \includegraphics[width=\linewidth]{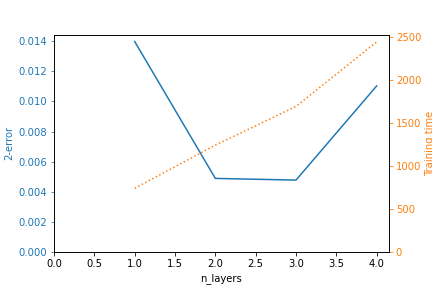}
        \caption{Linear scale}
    \end{subfigure}
    \begin{subfigure}{0.49\textwidth}
        \centering
        \includegraphics[width=\linewidth]{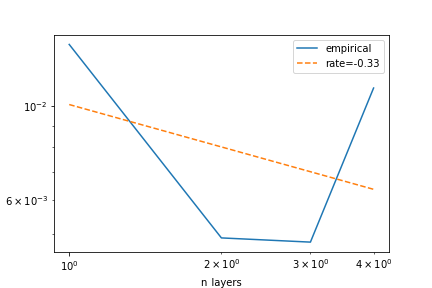}
        \caption{Logarithmic scale}
    \end{subfigure}
    \caption{$L^2$-error of the DGM for a call option in the Black--Scholes model against number of layers varying from 1 to 4.}
    \label{fig:n_layers_BS_DGM}
\end{figure}

\begin{figure}
    \centering
    \begin{subfigure}{0.49\textwidth}
        \centering
        \includegraphics[width=\linewidth]{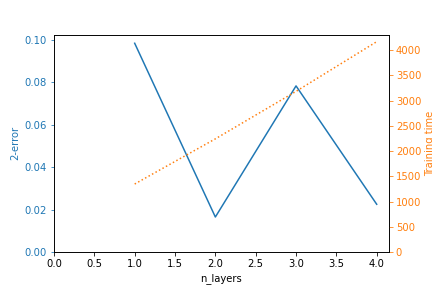}
        \caption{Linear scale}
    \end{subfigure}
    \begin{subfigure}{0.49\textwidth}
        \centering
        \includegraphics[width=\linewidth]{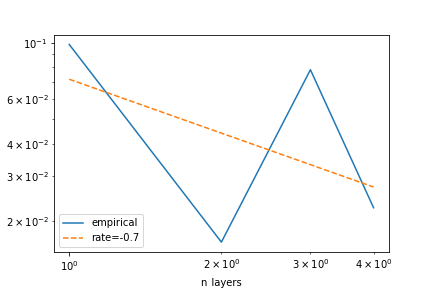}
        \caption{Logarithmic scale}
    \end{subfigure}
    \caption{$L^2$-error of the DGM for a call option in the Heston model against number of layers varying from 1 to 4.}
    \label{fig:n_layers_Heston_DGM}
\end{figure}

\begin{figure}
    \centering
    \begin{subfigure}{0.49\textwidth}
        \centering
        \includegraphics[width=\linewidth]{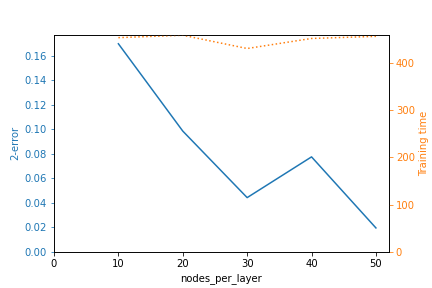}
        \caption{Linear scale}
    \end{subfigure}
    \begin{subfigure}{0.49\textwidth}
        \centering
        \includegraphics[width=\linewidth]{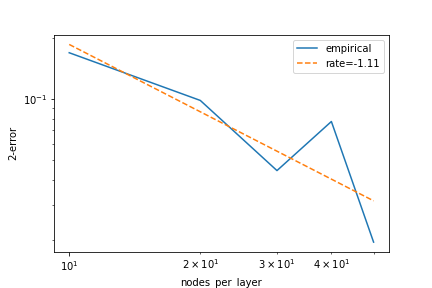}
        \caption{Logarithmic scale}
    \end{subfigure}
    \caption{$L^2$-error of the TDGF for a call option in the Black--Scholes model against number of nodes per layer varying from 10 to 50.}
    \label{fig:nodes_per_layer_BS_TDGF}
\end{figure}

\begin{figure}
    \centering
    \begin{subfigure}{0.49\textwidth}
        \centering
        \includegraphics[width=\linewidth]{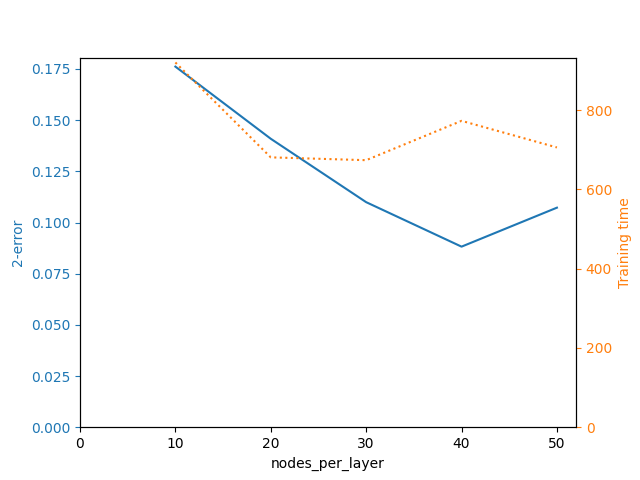}
        \caption{Linear scale}
    \end{subfigure}
    \begin{subfigure}{0.49\textwidth}
        \centering
        \includegraphics[width=\linewidth]{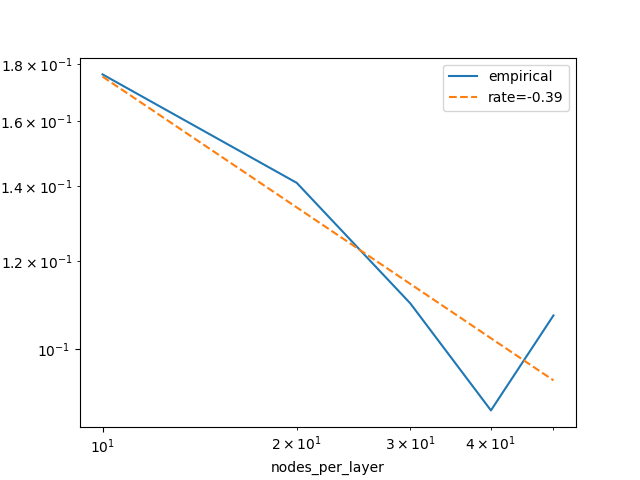}
        \caption{Logarithmic scale}
    \end{subfigure}
    \caption{$L^2$-error of the TDGF for a call option in the Heston model against number of nodes per layer varying from 10 to 50.}
    \label{fig:nodes_per_layer_Heston_TDGF}
\end{figure}

\begin{figure}
    \centering
    \begin{subfigure}{0.49\textwidth}
        \centering
        \includegraphics[width=\linewidth]{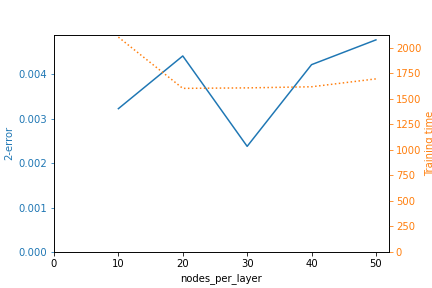}
        \caption{Linear scale}
    \end{subfigure}
    \begin{subfigure}{0.49\textwidth}
        \centering
        \includegraphics[width=\linewidth]{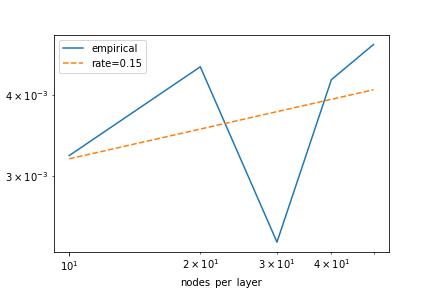}
        \caption{Logarithmic scale}
    \end{subfigure}
    \caption{$L^2$-error of the DGM for a call option in the Black--Scholes model against number of nodes per layer varying from 10 to 50.}
    \label{fig:nodes_per_layer_BS_DGM}
\end{figure}

\begin{figure}
    \centering
    \begin{subfigure}{0.49\textwidth}
        \centering
        \includegraphics[width=\linewidth]{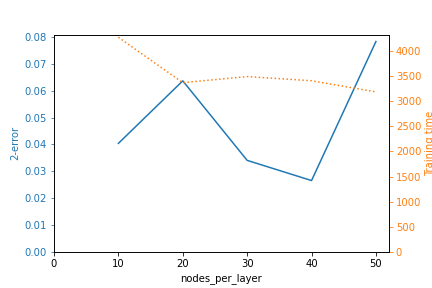}
        \caption{Linear scale}
    \end{subfigure}
    \begin{subfigure}{0.49\textwidth}
        \centering
        \includegraphics[width=\linewidth]{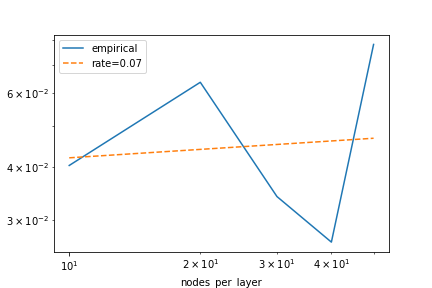}
        \caption{Logarithmic scale}
    \end{subfigure}
    \caption{$L^2$-error of the DGM for a call option in the Heston model against number of nodes per layer varying from 10 to 50.}
    \label{fig:nodes_per_layer_Heston_DGM}
\end{figure}

\begin{figure}
    \centering
    \begin{subfigure}{0.49\textwidth}
        \centering
        \includegraphics[width=\linewidth]{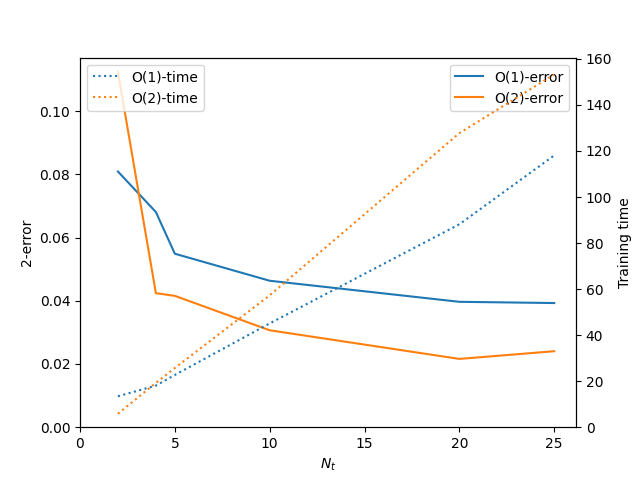}
        \caption{Linear scale}
    \end{subfigure}
    \begin{subfigure}{0.49\textwidth}
        \centering
        \includegraphics[width=\linewidth]{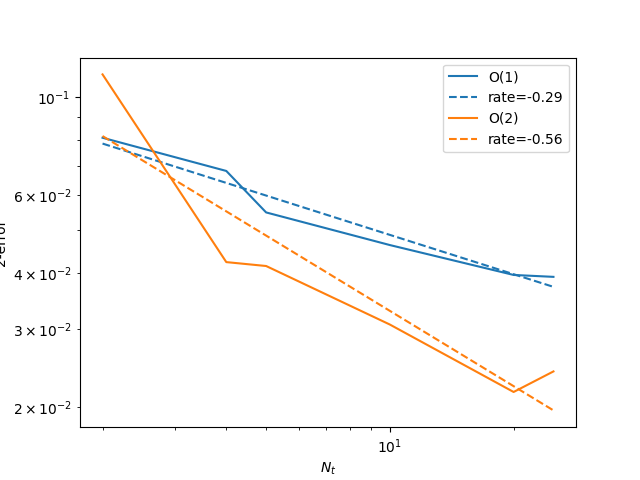}
        \caption{Logarithmic scale}
    \end{subfigure}
    \caption{$L^2$-error of the TDGF for a call option in the Black--Scholes model against number of time steps varying from 2 to 25.}
    \label{fig:N_t_BS_TDGF}
\end{figure}

\begin{figure}
    \centering
    \begin{subfigure}{0.49\textwidth}
        \centering
        \includegraphics[width=\linewidth]{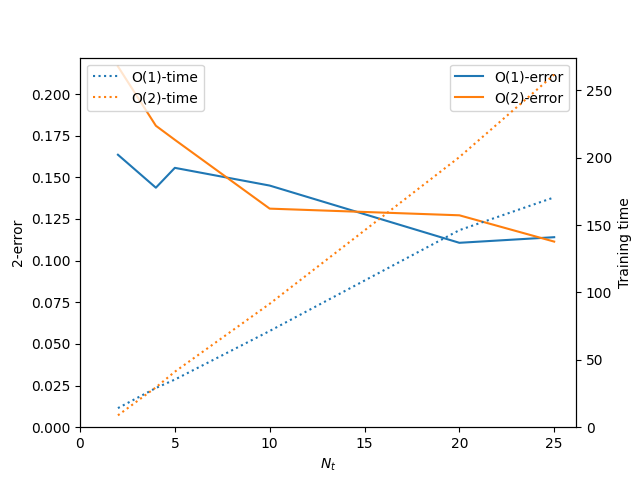}
        \caption{Linear scale}
    \end{subfigure}
    \begin{subfigure}{0.49\textwidth}
        \centering
        \includegraphics[width=\linewidth]{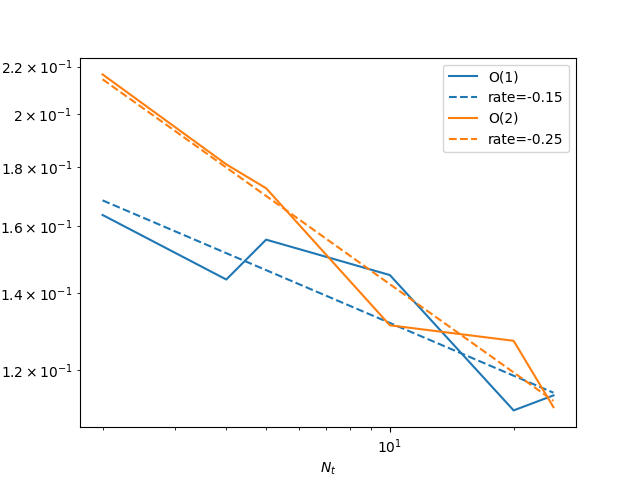}
        \caption{Logarithmic scale}
    \end{subfigure}
    \caption{$L^2$-error of the TDGF for a call option in the Heston model against number of time steps varying from 2 to 25.}
    \label{fig:N_t_Heston_TDGF}
\end{figure}

\end{document}